\documentclass[aps,prb,twocolumn,superscriptaddress,showpacs,floatfix]{revtex4}
\usepackage{bm}
\usepackage{amssymb}
\usepackage{graphicx,epsfig}

\newcommand{\nn}{\nonumber \\}

\def\da     {\downarrow}
\def\ua     {\uparrow}

\begin{document}
\title{Implementation of three-qubit Toffoli gate in a single step }
\author{Ai Min \surname{Chen}}
\author{Sam Young \surname{Cho}}
\affiliation{Centre for Modern Physics and Department of Physics,
Chongqing University, Chongqing 400044, The People's Republic of China}
\author{Mun Dae \surname{Kim}}
\email[Corresponding author;E-mail address:]{mdkim@yonsei.ac.kr}
\affiliation{Institute of Physics and Applied Physics, Yonsei University, Seoul 120-749, Korea }
\affiliation{Korea Institute for Advanced Study, Seoul 130-722, Korea}

\begin{abstract}

Single-step implementations of multi-qubit gates are generally
believed  to provide a simpler design, a faster operation, and a
lower decoherence. For coupled three qubits interacting with a
photon field, a realizable scheme for a single-step Toffoli gate
is investigated. We find that the three qubit system can be
described by four effective modified Jaynes-Cummings models in the
states of two control qubits. Within the rotating wave
approximation, the modified Jaynes-Cummings models are shown to be
reduced to the conventional Jaynes-Cummings models with
renormalized couplings between qubits and photon fields. A
single-step Toffoli gate is shown to be realizable with tuning the
four characteristic oscillation periods that satisfy a
commensurate condition. Possible values of system parameters are
estimated for single-step Toffli gate. From numerical calculation,
further, our single-step Toffoli gate operation errors are
discussed due to imperfections in system parameters, which shows
that a Toffoli gate with high fidelity can be obtained by
adjusting pairs of the photon-qubit and the qubit-qubit coupling
strengthes. In addition, a decoherence effect on the Toffoli gate
operation is discussed due to a thermal reservoir.

\end{abstract}

\maketitle

\section{Introduction}

Any quantum computing algorithm can be decomposed into
a sequence of single qubit rotations and entangling two-qubit gates. \cite{universal}
Although this universal gate provides a convenient and intuitive design
for quantum algorithm, the design becomes so complicated
as the qubits scale up that it may be hard to implement the quantum algorithm.
The quantum gates with more than two qubits, thus, will be very
useful for constructing  quantum computing algorithms and quantum
error correction protocols.
Physical realization of the multi-qubit gate
has been intensively studied with various qubit systems.

Among the multi-qubit gate,  the Toffoli gate \cite{Toffoli} (Controlled-Controlled-NOT or C$^2$-NOT gate)
provides a universal gate set for classical computing.
Recently, the three-qubit quantum Toffoli gate has been
achieved, for example, with NMR qubits \cite{Cory,Cory2},
ion-trap qubits, \cite{Monz} and superconducting circuit. \cite{Wallraff}
The Toffoli gate  has usually been implemented by decomposing
it into a sequence of single qubit operations and CNOT gates.
However, the single-step implementation  \cite{Grig} of multi-qubit
gate will make the circuit design simpler and provide shorter gate operation time.
Further, it may achieve the lowest possible qubit decoherence,
providing high fidelity.
For the atoms in cavity the inter-atom interaction is mediated by the photon mode.
In this case the schemes for the single-step multi-qubit controlled-phase-flip gate \cite{Lin}
and Toffoli gate \cite{Duan}  have been proposed.
In this study, we analyze a scheme for realizing
the {\it single-step} implementation of Toffoli gate
for solid-state qubits with Ising-type interaction.

We consider a system of three qubits coupled by
Ising interaction, where the qubit state evolution is driven by an external
oscillating field. Two of the qubits behave as the
control qubit, while the third one  as the target.
The external photon field is resonant with the target qubit
for a specific control qubit state,
while it is off-resonant for the other control qubit states.
For the resonant case, the Hamiltonian is described by the Jaynes-Cummings model
with a  photon field, coupled vertically with the qubit state in a rotated coordinate.
However, for the off-resonant case, the photon field has components parallel
as well as vertical to the qubit basis.
In this paper, we analyze this  {\it modified Jaynes-Cummings model}
to find the conditions for the single-step Toffoli gate.
These conditions  can be obtained by
investigating the commensurate oscillation periods of the target
qubit states. The gate operation error is estimated to be so small
that our scheme has advantages over the decomposition scheme of Toffoli gate.
The decoherence effect of the environment is analyzed by introducing
the interaction between the qubit and the thermal reservoir.

\section{Modified Jaynes-Cummings model}

\subsection{Effective Hamiltonian}

A qubit  interacting with a microwave field, $g\cos\omega t$,
with coupling constant $g$ can be described by a  semiclassical
Hamiltonian
\begin{eqnarray}
\label{Hone}
H^{\rm sc}_{\rm qubit}=\left(\frac{\epsilon}{2}+g\cos\omega t\right)\sigma_z- t_q\sigma_x,
\end{eqnarray}
where $\epsilon$ is the qubit energy splitting,
$t_{q}$ is the tunnelling amplitude between different (pseudo-)
spin states, and $\sigma_{z,x}$ are the Pauli matrices.
The microwave field can be quantized into photon field and, if we
consider only the single-photon process, the system can be described by the Hamiltonian,
\begin{eqnarray}
H_{\rm qubit}=\hbar\omega a^\dagger a+\frac{\epsilon+g(a+a^\dagger)}{2}\sigma_z- t_q\sigma_x,
\label{oneH}
\end{eqnarray}
where  $a$ and $a^\dagger$ are the photon annihilation and
creation operators.
This type of coupling can be seen, for example, in the circuit-QED architecture \cite{Blais,Blais2,Blais3} and the
superconducting flux qubit. \cite{Mooij,Mooij2,Mooij3,Mooij4}
For three coupled qubits (qubits A,B,C) the Hamiltonian in the basis of $\{|\da\rangle, |\ua\rangle\}$
is given by
\begin{eqnarray}
\label{H}
H\!\!\!\!&=&\!\!\!\!\frac{E_A}{2}\sigma_z\!\!\otimes\! \sigma_0\!\!\otimes\! \sigma_0\!\!
+\!\!\frac{E_B}{2} \sigma_0\!\!\otimes\! \sigma_z\!\!\otimes\! \sigma_0
\!\!+\!\!\frac{E_C}{2} \sigma_0\!\!\otimes\! \sigma_0\!\!\otimes\! \sigma_z\\
&&\!\!\!\!-t^A_q \sigma_x\!\!\otimes\! \sigma_0\!\!\otimes\!  \sigma_0\!-\!t^B_q\sigma_0\!\!\otimes\! \sigma_x \!\!\otimes\! \sigma_0
\!-\!t^C_q\sigma_0\!\!\otimes\! \sigma_0 \!\!\otimes\! \sigma_x \nn
&&\!\!\!\!+J^{AB}\sigma_z\!\!\otimes\!\sigma_z\!\!\otimes\! \sigma_0\!+\!J^{AC}\sigma_z\!\!\otimes\! \sigma_0\!\!\otimes\!\sigma_z
\!+\!J^{BC}\sigma_0\!\!\otimes\! \sigma_z\!\!\otimes\!\sigma_z\nonumber.
\end{eqnarray}
Here $E_i=\epsilon_i+g(a+a^\dagger)$ with $i=A,B,C$, and
the Ising-type couplings are set to be equal to each other,
$J^{AB}=J^{AC}=J^{BC}=J$.

\begin{figure}[b]
\vspace{8cm}
\includegraphics{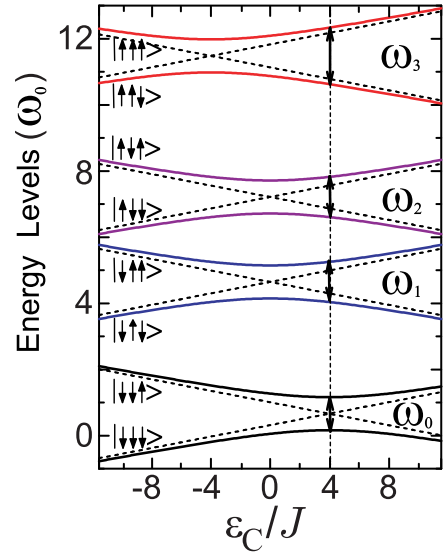}
\vspace{-0.1cm}
\caption{(Color online) Dotted lines are the energy levels $E_{ss's''}$, and
the solid lines correspond to the energy levels,
$(E_{ss'\uparrow}+E_{ss'\downarrow})/2\pm (1/2)\hbar\omega_j$, with $j=0,1,2,3$.
The operating point is $\epsilon_C=4J$.}
\label{fig1}
\end{figure}

In Fig. \ref{fig1} the diagonal energies  $E_{ss's''}$ ($s\in \{\downarrow,\uparrow\}$) of this Hamiltonian,
excluding the qubit-photon interaction term $g(a+a^\dagger)$, are
plotted as dotted lines. We here set the  first two  qubits
as the control qubits while  the third qubit as the target qubit.
We consider the parameter regime, $\epsilon_A, \epsilon_B \gg J$,
so that the energy levels of coupled qubits $|\downarrow\downarrow s''\rangle$
are far lower while those of  $|\uparrow\uparrow s''\rangle$ much higher
than the energy levels of $|s,-ss''\rangle$ as shown in  Fig. \ref{fig1}.
Here, $-s$ is opposite spin of $s$.
Then, the tunnellings between the states $|sss''\rangle$ and  $|s,-ss''\rangle$
are suppressed in the Hamiltonian; thus, we can set $t^A_q=0$ and $t^B_q=0$.
Hence, the three qubit Hamiltonian of Eq. (\ref{H}) can be represented
as a block-diagonal form,
\begin{equation}
H=H_{|\uparrow\uparrow\rangle}\oplus H_{|\uparrow\downarrow\rangle}\oplus
H_{|\downarrow\uparrow\rangle}\oplus H_{|\downarrow\downarrow\rangle},
\end{equation}
where $H_{|ss'\rangle}$ is explicitly given as
\begin{eqnarray}
H_{|ss'\rangle}&=&\hbar\omega a^\dagger a +\frac{E_{ss'\uparrow}+E_{ss'\downarrow}}{2}\sigma_0 \nonumber\\
&&+\frac{E_{ss'\uparrow}-E_{ss'\downarrow}+g(a+a^\dagger)}{2}\sigma_z-t^C_q\sigma_x,
\label{twoH}
\end{eqnarray}
and $|ss'\rangle$ indicates the control qubit states.

In order to analyze the Rabi oscillation of this system
we diagonalize the qubit part of the Hamiltonian by introducing a coordinate transformation as follows,
\begin{eqnarray}
V=\exp\bigg(-\frac{i}{2} \sigma_y
\theta_{\uparrow\uparrow}\bigg)\oplus
\exp\bigg(-\frac{i}{2} \sigma_y
\theta_{\uparrow\downarrow}\bigg)\nonumber\\
\oplus \exp\bigg(-\frac{i}{2} \sigma_y
\theta_{\downarrow\uparrow}\bigg)\oplus
\exp\bigg(-\frac{i}{2} \sigma_y
\theta_{\downarrow\downarrow}\bigg)
\label{V}
\end{eqnarray}
with
\begin{eqnarray}
\tan\theta_{j}=\frac{2t^C_q}{|E_{ss'\uparrow}-E_{ss'\downarrow}|}.
\end{eqnarray}
Here, note that we assign $j=0,1,2,3$ for
$|ss'\rangle=|\downarrow\downarrow\rangle,
|\downarrow\uparrow\rangle, |\uparrow\downarrow\rangle,
|\uparrow\uparrow\rangle$, respectively. Then, the transformed
Hamiltonian ${\cal H}=V^{-1}HV$ is given by
\begin{eqnarray}
{\cal H}={\cal H}_{|\uparrow\uparrow\rangle}\oplus {\cal H}_{|\uparrow\downarrow\rangle}\oplus
{\cal H}_{|\downarrow\uparrow\rangle}\oplus {\cal H}_{|\downarrow\downarrow\rangle}.
\end{eqnarray}
Among the energy levels in Fig. \ref{fig1}, the lowest two energy levels are given as
$E_{\downarrow\downarrow\downarrow}=-(\epsilon_A+\epsilon_B+\epsilon_C)/2+3J$
and  $E_{\downarrow\downarrow\uparrow}=-(\epsilon_A+\epsilon_B-\epsilon_C)/2-J$
from the Hamiltonian in Eq. (\ref{H}).
These two energy levels are degenerate when $\epsilon_C=4J$, and
we will use this degeneracy point as the operating point.

In general, the Hamiltonian  is represented as
a modified Jaynes-Cummings model as follows,
\begin{eqnarray}
\label{H1}
{\cal H}_{|ss'\rangle} &=& \hbar\omega a^\dagger a + \frac12[\hbar\omega_j +
\beta_j g(a+a^\dagger)]\sigma_z \nonumber\\
&&+\frac{\alpha_j g}{2}(a+a^\dagger)(\sigma_+ +\sigma_-),
\end{eqnarray}
where
\begin{eqnarray}
\alpha_j=\sin\theta_{j}, ~~\beta_j=\cos\theta_{j},\\
\hbar\omega_j=\sqrt{\Delta E_{ss'}^2+(2t^C_q)^2}
\end{eqnarray}
with $\Delta E_{ss'}=E_{ss'\uparrow}-E_{ss'\downarrow}$, and we drop the
irrelevant term $g(a+a^\dagger)\sigma_0$. This Hamiltonian has the
photon field which has the components  either vertical or parallel
to the qubit with coupling constant $\alpha_j g$ or $\beta_j g$.

In the Hamiltonian of Eq. (\ref{H1}) the Pauli operators are represented in
the basis of $\{|ss'g\rangle,|ss'e\rangle\}$, where
\begin{eqnarray}
|ss'g\rangle \!\!&=&\!\! \cos(\theta_{j}/2)|ss'\downarrow\rangle+\sin(\theta_{j}/2)|ss'\uparrow\rangle, \nonumber\\
|ss'e\rangle \!\!&=&\!\! -\sin(\theta_{j}/2)|ss'\downarrow\rangle+\cos(\theta_{j}/2)|ss'\uparrow\rangle,
\label{ssge}
\end{eqnarray}
whose energy levels are shown as solid lines in Fig. \ref{fig1}.
Here, $|ss'\rangle$ denotes the control qubit state, and
$|g\rangle$ and $|e\rangle$ denote the ground and excited states
of the target qubit state.
At the operating point  $\epsilon_C=4J$,
\begin{eqnarray}
\Delta E_{ss'}=0, 4J, 4J, ~{\rm and} ~8J
\end{eqnarray}
for $|ss'\rangle=|\downarrow\downarrow\rangle, |\downarrow\uparrow\rangle,
|\uparrow\downarrow\rangle$, and $|\uparrow\uparrow\rangle$, respectively.
Since the Hamiltonian ${\cal H}_{|\downarrow\downarrow\rangle}$
has degeneracy at this operating point,
${\cal H}_{|\downarrow\downarrow\rangle}$ is reduced
to the usual Jaynes-Cummings model, \cite{Jaynes}
\begin{eqnarray}
{\cal H}_{|\downarrow\downarrow\rangle}=\hbar\omega a^\dagger a + \frac{\hbar\omega_0}{2}\sigma_z
+\frac{g}{2}(a+a^\dagger)(\sigma_+ +\sigma_-)
\end{eqnarray}
with
\begin{eqnarray}
\hbar\omega_0=2t^C_q.
\end{eqnarray}

In order to describe the dynamics of qubit system,
we introduce a rotating coordinate such as $\psi(t)=U(t)\phi(t)$, where
\begin{eqnarray}
U(t)\!=\exp\left[-it\left(\frac{1}{2}\hbar\omega_0 \sigma_z
+\hbar\omega a^\dagger a  \right)\right].
\end{eqnarray}
Accordingly,  the Schr{\" o}dinger equation
${\cal H}\psi(t)=i\hbar \frac{\partial}{\partial t} \psi(t)$
is written as $i \hbar\frac{\partial}{\partial t} \phi(t)={\cal H}^{\rm I} \phi (t)$
with ${\cal H}^{\rm I}=U^{-1}(t){\cal H}U(t)-i\hbar U^{-1}(t)(dU(t)/dt)$.
In this interaction picture, the Hamiltonian ${\cal H}^{\rm I}_{|ss'\rangle}$ is represented as
\begin{eqnarray}
{\cal H}^{\rm I}_{|ss'\rangle} =  \frac12[\hbar(\omega_j-\omega_0) +
\beta_j g(a(t)+a^\dagger(t))]\sigma_z \nonumber\\
+\frac12\alpha_j g(a(t)+a^\dagger(t))(\sigma_+(t) +\sigma_-(t)),
\end{eqnarray}
where the transformed operators ${\cal O}(t)=U^{-1}(t){\cal O} U(t)$ are given by
\begin{eqnarray}
\label{sigma}
\sigma_\pm(t)=\sigma_\pm e^{\pm i\omega_0t},
~a^\dagger(t)=a^\dagger e^{i\omega t},
~a(t)=a e^{-i\omega t}.
\end{eqnarray}

The parallel coupling term between the qubit and the photon field,
$\beta_j g(a(t)+a^\dagger(t))\sigma_z$, in the  Hamiltonian ${\cal H}^{\rm I}_{|ss'\rangle}$
can be eliminated
by introducing another coordinate  transformation,
\begin{eqnarray}
W(t)=\exp\left[-\frac{\beta_jg}{2\hbar\omega_0}(a^\dagger(t)-a(t))\sigma_z\right].
\end{eqnarray}

The transformed Hamiltonian
${\tilde {\cal H}}^{\rm I}_{|ss'\rangle}
=W^{-1}(t){\cal H}^{\rm I}_{|ss'\rangle} W(t)-i\hbar W^{-1}(t)(dW(t)/dt)$ is then written as
\begin{eqnarray}
{\tilde {\cal H}}^{\rm I}_{|ss'\rangle} &=&  \frac12[\hbar(\omega_j-\omega_0) +
\beta_j g({\tilde a}(t)+{\tilde a}^\dagger(t))]\sigma_z \nonumber\\
&&+\frac12\alpha_j g({\tilde a}(t)+{\tilde a}^\dagger(t))
({\tilde \sigma}_+(t) +{\tilde \sigma}_-(t))\nonumber\\
&&-\frac12\beta_j g(a(t)+a^\dagger(t))\sigma_z.
\end{eqnarray}
For photon field the transformed operators  ${\tilde {\cal O}}(t)=W^{-1}{\cal O}(t)W(t)$
are represented as \cite{Vogel}
\begin{eqnarray}
{\tilde a}(t)&=&a(t)-(\beta_jg/2\hbar\omega_0)\sigma_z, \nonumber\\
{\tilde a}^\dagger(t)&=&a^\dagger(t)-(\beta_jg/2\hbar\omega_0)\sigma_z,
\end{eqnarray}
and for spin operators \cite{Sakurai}
\begin{eqnarray}
{\tilde \sigma}_\pm(t)=\sigma_\pm(t)\exp[\pm(\beta_jg/\hbar\omega_0)(a^\dagger(t)-a(t))].
\end{eqnarray}
Then, the Hamiltonian becomes
\begin{eqnarray}
\label{tHI}
{\tilde {\cal H}}^{\rm I}_{|ss'\rangle} &=&  \frac12\hbar(\omega_j-\omega_0)\sigma_z
+\frac12 \alpha_j g\left(a(t)+a^\dagger(t)-\frac{\beta_jg}{\hbar\omega_0}\sigma_z\right)\nonumber\\
&&\times\left(\sigma_+(t)\exp[\frac{\beta_jg}{\hbar\omega_0}(a^\dagger(t)-a(t))]\right.\nonumber\\
&&\left.+\sigma_-(t)\exp[-\frac{\beta_jg}{\hbar\omega_0}(a^\dagger(t)-a(t))]\right),
\end{eqnarray}
dropping a constant term.

The matrix element of the Hamiltonian can be evaluated
in the rotating wave approximation  (RWA) \cite{Jaynes}
as follows, [see Appendix]
\begin{eqnarray}
\label{off1}
&&\langle g,n+1|\left(a(t)+a^\dagger(t)-\gamma_j\sigma_z\right)e^{-\gamma_j(a^\dagger(t)-a(t))}
\sigma_-(t)|e,n\rangle \nonumber\\
&&=\left(\sqrt{n+1}L^0_n(\gamma_j^2)+\frac{\gamma_j^2}{\sqrt{n+1}}[L^2_n(\gamma_j^2)-L^1_n(\gamma_j^2)]\right)
\nonumber\\ && ~~~\times e^{-\gamma_j^2/2}\equiv F_n(\gamma_j), \\
\label{off2}
&&\langle e,n|\left(a(t)+a^\dagger(t)-\gamma_j\sigma_z\right)e^{\gamma_j(a^\dagger(t)-a(t))}
\sigma_+(t)|g,n+1\rangle \nonumber\\
&&=\left(\sqrt{n+1}L^0_{n+1}(\gamma_j^2)+\frac{\gamma_j^2}{\sqrt{n+1}}[L^2_{n-1}(\gamma_j^2)+L^1_n(\gamma_j^2)]\right)
\nonumber\\ &&~~~\times e^{-\gamma_j^2/2}\equiv G_n(\gamma_j).
\end{eqnarray}
We can check that those two matrix elements are equivalent with each other,
\begin{eqnarray}
 F_n(\gamma_j)=G_n(\gamma_j),
\end{eqnarray}
by using  the recurrent relations, \cite{Abra}
\begin{eqnarray}
&&L^2_n(\gamma_j^2)-L^1_n(\gamma_j^2)=L^2_{n-1}(\gamma_j^2), \\
&&(n+1)L^0_{n+1}(\gamma_j^2)+\gamma_j^2L^1_n(\gamma_j^2)= (n+1)L^0_n(\gamma_j^2). \nonumber
\end{eqnarray}
As a result, in the RWA  the modified Jaynes-Cummings model is reduced to the
conventional Jaynes-Cummings model with a renormalized coupling between the qubit and
photon field.
Hence, the effective Hamiltonian in the RWA reads
\begin{eqnarray}
\label{HRWA}
{\tilde {\cal H}}^{\rm I,RWA}_{|ss'\rangle}=\frac12 \hbar(\omega_j-\omega_0)\sigma_z
+\frac{\tilde{g}_{jn}}{2}(a \sigma_++a^\dagger \sigma_-)
\end{eqnarray}
in the basis of $\{|e,n\rangle_j, |g,n+1\rangle_j\}$, where
\begin{eqnarray}
\tilde{g}_{jn}=\alpha_jg F_n(\gamma_j)
\end{eqnarray}
is the renormalized coupling constant.

\begin{table*}[hbtp!]
\vspace{0cm}
\centerline{
\begin{tabular}{c c c c c c c c c c c}
\hline
\hline
 $(p,q)$&~$J/t^C_q$&~$g/t^C_q$&~$\alpha_1$&~$\alpha_3$&~$\beta_1$&~$\beta_3$&~$\gamma_1$&~$\gamma_3$ & ~${\tilde g}_{10}/g$ &~${\tilde g}_{30}/g$ \\
\hline
(1,3)    &  ~0.340 (0.358) &  ~0.230 (0.254) & ~0.822 & ~0.585 &~0.569  &~0.811 & ~0.067 & ~0.096 & 0.820 & 0.582 \\
(2,6)    &  ~0.460 (0.484) &  ~0.182 (0.199) & ~0.736 & ~0.478 &~0.677  &~0.879 & ~0.062 & ~0.080 & 0.734 & 0.476\\
(3,9)    &  ~0.477 (0.471) &  ~0.128 (0.126) & ~0.724 & ~0.464 &~0.690  &~0.886 & ~0.044 & ~0.057 & 0.723 & 0.463\\
\hline
\hline
\end{tabular}
}
\vspace{-0.cm}
\caption{ The values of $J$ and $g$ are obtained by solving Eq. (\ref{condition}).
Here the values in parenthesis are obtained from numerical calculation,
which fit well with the analytic results for small $\gamma$ where
the RWA works well. $\alpha_j$($\beta_j$) corresponds to the vertical (parallel) coupling
strength between the qubit and photon field. ${\tilde g}_{j0}$ is the renormalized coupling.}
\label{table}
\end{table*}

\subsection{Commensurate Condition}

In this paper, we are concentrated on the ground state with $n=0$.
Then, the coupling constant ${\tilde g}_{j0}= \alpha_jg e^{-\frac12\gamma_j^2}$
is Gaussian. Though  the coupling constant obtained by semiclassical analysis
in Ref. \onlinecite{comm} is oscillatory, it fits well with the present result
for small $\gamma_j$ because the RWA works well for a weak coupling.
%
For $n=0$, the Hamiltonian in Eq. (\ref{HRWA}) describes the quantum oscillation
between the states $|g,1\rangle_j$ and $|e,0\rangle_j$
with the frequency,
\begin{eqnarray}
\Omega_j = \sqrt{\left(\omega_j-\omega_0\right)^2
+\left({\tilde g}_{j0}/\hbar\right)^2}.
\label{Omega1}
\end{eqnarray}
$\Omega_j$ is the oscillation frequency between the two states,
 $|g,1\rangle_j=|ss'g\rangle$ and
$|e,0\rangle_j=|ss'e\rangle$.

In this study, our Hamiltonian in the RWA is written as
\begin{eqnarray}
{\tilde {\cal H}}^{\rm I,RWA}=
{\tilde {\cal H}}^{\rm I,RWA}_{|\uparrow\uparrow\rangle} \oplus
{\tilde {\cal H}}^{\rm I,RWA}_{|\uparrow\downarrow\rangle} \oplus
{\tilde {\cal H}}^{\rm I,RWA}_{|\downarrow\uparrow\rangle} \oplus
{\tilde {\cal H}}^{\rm I,RWA}_{|\downarrow\downarrow\rangle},
\end{eqnarray}
and we want to flip the target qubit state when the control qubit state
is $|\downarrow\downarrow\rangle$.
For $j=0$, the Hamiltonian of Eq. (\ref{HRWA}) is reduced to
${\tilde {\cal H}}^{\rm I,RWA}_{|\downarrow\downarrow\rangle}=\frac{g}{2}(a \sigma_++a^\dagger \sigma_-),$
which describes the Rabi oscillation between the states, $|\downarrow\downarrow g\rangle$ and
$|\downarrow\downarrow e\rangle$ with the Rabi frequency
\begin{eqnarray}
\Omega_{\rm R}=\Omega_0=g/\hbar.
\end{eqnarray}
On the other hand, for $j \neq 0$  the Hamiltonian demonstrates  a non-Rabi oscillation
between the states, $|ss' g\rangle$ and $|ss' e\rangle$ with the oscillating frequency $\Omega_j$.
The Toffoli gate requires that for a specific control qubit state the target qubit state
flips, while for the other control qubit states the target qubit state remains at the original state.
The Toffoli gate  is represented as following matrix,
\begin{eqnarray}
M_{\rm Toffoli}=
\left[\matrix{ 1 & 0 & 0 & 0 & 0 & 0 & 0 & 0 \cr
 0 & 1 & 0 & 0 & 0 & 0 & 0 & 0 \cr
 0 & 0 & 1 & 0 & 0 & 0 & 0 & 0 \cr
 0 & 0 & 0 & 1 & 0 & 0 & 0 & 0 \cr
 0 & 0 & 0 & 0 & 1 & 0 & 0 & 0 \cr
 0 & 0 & 0 & 0 & 0 & 1 & 0 & 0 \cr
0 & 0 & 0 & 0 & 0 & 0 & 0 & 1 \cr
0 & 0 & 0 & 0 & 0 & 0 & 1 & 0 }\right]
\end{eqnarray}
in the basis of $\{|\uparrow\uparrow e\rangle, |\uparrow\uparrow g\rangle,
|\uparrow\downarrow e\rangle, |\uparrow\downarrow g\rangle, |\downarrow\uparrow e\rangle,
|\downarrow\uparrow g\rangle, |\downarrow\downarrow e\rangle, |\downarrow\downarrow g\rangle \}$.

Let us consider the initial state,  $|\Psi(0)\rangle=(1/2)(|\downarrow\downarrow g\rangle
+|\downarrow\uparrow g\rangle+|\uparrow\downarrow g\rangle+|\uparrow\uparrow g\rangle)$.
When $t=T_0/2$ with $T_0=2\pi/\Omega_0$, the  state
$|\downarrow\downarrow g\rangle$ flips to the state $|\downarrow\downarrow e\rangle$,
while the other stats  $|ss' g\rangle$ also evolve during the time.
Hence, in general, the Toffoli cannot be achieved at time $t=T_0/2$.
However, if half the oscillation period, $\pi/\Omega_0$, is
an integer multiple of the other oscillation periods, $2\pi/\Omega_j$ with $j\neq 0$,
as follows,
\begin{eqnarray}
\Omega_1&=&\Omega_2=2p\Omega_0,\nonumber\\
\Omega_3&=&2q\Omega_0
\label{condition}
\end{eqnarray}
with integers $p, q$, the Toffoli gate can be achieved.
Here, since $\Delta E_{\uparrow\downarrow}=E_{\downarrow\uparrow}=4J$,
$\omega_1=\omega_2$, $\alpha_1=\alpha_2$, and  $\beta_1=\beta_2$, we have
$\Omega_1=\Omega_2$.
At time $t_k=(2k+1)\pi/\Omega_0$ with integer $k$, the qubit states with $j=0$ flips to
the other state whereas the states with $j=1,2,3$ come back to the original states.
Consequently, if $\Omega_j$'s satisfy the conditions in Eq. (\ref{condition}),
the Toffoli gate can be achieved at time $t_k$.


The coupled equations of Eq. (\ref{condition}) for commensurate condition  can be solved numerically,
and partial results are  summarized in Table \ref{table}. The coupled equations determine the values of
$J/t^C_q$ and $g/t^C_q$ for a given pair of $(p,q)$. Considering that the typical tunnelling frequency
of solid-state qubits is 1-2 GHz, the interqubit coupling strength $J/h$ and the Rabi frequency $g/h$
should be of the order of 100MHz. The values of $\alpha_j$'s and $\beta_j$'s
are the prefactor of the coupling vertical and parallel to the qubit states, respectively,
in the Hamiltonian of Eq. (\ref{H1}).
${\tilde g}_{j0}$ is the  renormalized coupling constant.
We found that if $q=3p$ there are solutions for the coupled equation of Eq. (\ref{condition}),
but we also found that there exist solutions for some pairs of $(p,q)$ around the line, $q=3p$.

\section{Numerical Analysis of Toffoli gate}

If the conditions for commensurate oscillation in Eq. (\ref{condition})
are satisfied, the analysis of previous section provides a complete (maximum fidelity)
Toffoli gate operation. However, the analysis neglected the multi-photon processes and
was performed under the RWA. In reality, thus, the gate operation cannot be complete.
Through a numerical analysis with the original semi-classical
Hamiltonian, we can estimate the error of our Toffoli gate operation scheme.

Similarly to the argument of section II, the three qubit Hamiltonian can be written
as following block-diagonal form in a semiclassical description
\begin{equation}
H^{\rm sc}=H^{\rm sc}_{|\uparrow\uparrow\rangle}\oplus H^{\rm sc}_{|\uparrow\downarrow\rangle}\oplus
H^{\rm sc}_{|\downarrow\uparrow\rangle}\oplus H^{\rm sc}_{|\downarrow\downarrow\rangle},
\label{Hsc}
\end{equation}
where $H^{\rm sc}_{|ss'\rangle}=H^{\rm qubit}_{|ss'\rangle}+H^{\rm mw}_{|ss'\rangle}$ with
\begin{eqnarray}
H^{\rm qubit}_{|ss'\rangle}&=&\frac{E_{ss'\uparrow}-E_{ss'\downarrow}}{2}\sigma_z-t^C_q\sigma_x
+\frac{E_{ss'\uparrow}+E_{ss'\downarrow}}{2}\sigma_0,\nonumber\\
H^{\rm mw}_{|ss'\rangle}&=&g\cos\omega t\sigma_z.
\label{scTwo}
\end{eqnarray}
The eigenstates of $H^{\rm qubit}_{|ss'\rangle}$ are given by the states $|ss'g\rangle$ and
$|ss'e\rangle$ in Eq. (\ref{ssge}) whose evolutions are driven by $H^{\rm mw}_{|ss'\rangle}$.
Here we adjust the operating point as $\epsilon_C=4J$ and
set the initial states as $|\Psi(0)\rangle=(1/2)(|\downarrow\downarrow g\rangle
+|\downarrow\uparrow g\rangle+|\uparrow\downarrow g\rangle+|\uparrow\uparrow g\rangle)$ as before.

\begin{figure}[b]
\vspace{4.2cm}
\includegraphics{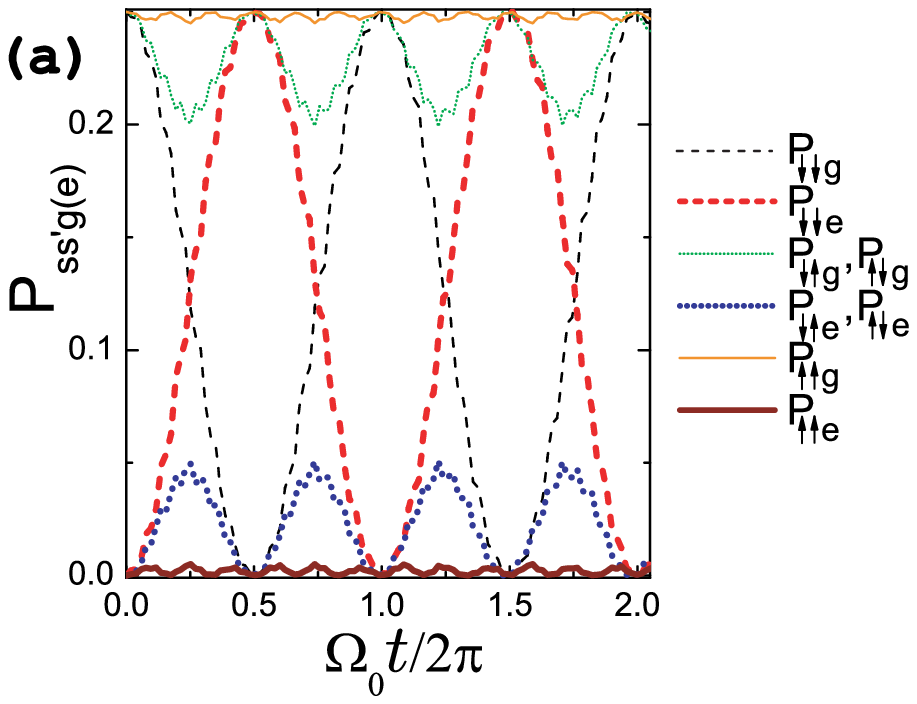}
\vspace{4.5cm}
\includegraphics{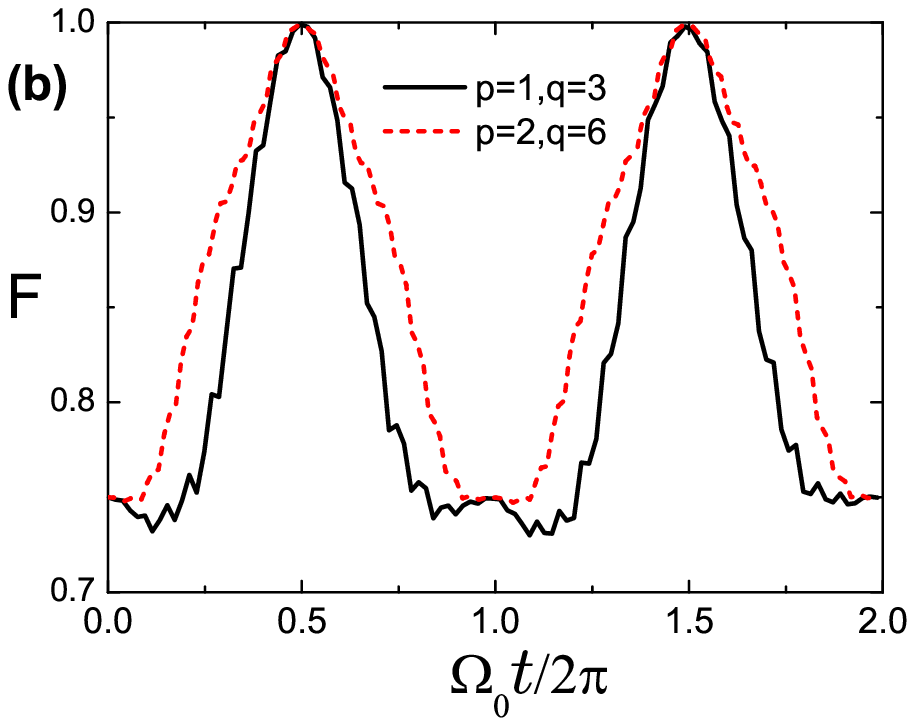}
\vspace*{1.2cm}
\caption{(Color online) (a) Time evolution of the initial state $|\Psi(0)\rangle$
by using the values of $J$ and $g$ obtained by the RWA  with $(p,q)$=(1,3)
in Table \ref{table}.  At $\Omega t=\pi$, the state $|\downarrow\downarrow g\rangle$ flips to
the state $|\downarrow\downarrow e\rangle$, while the other states come back to their initial
state. As a result, the Toffoli gate is achieved in a single step.
(b) Fidelity of the Toffoli gate  for $(p,q)$=(1,3) (solid line).
The fidelity approaches local maxima at $\Omega_0 t=(2k+1)\pi$.
Dotted line shows the fidelity for $(p,q)$=(2,6). }
\label{osc}
\end{figure}

The Toffoli gate can be demonstrated by using the values of $J$ and $g$
obtained analytically in Table \ref{table}.
Here, we use  the values for $(p,q)=(1,3)$.
In Fig.  \ref{osc} (a) we can observe that at $\Omega_0t=\pi$ the initial state
$|\downarrow\downarrow g\rangle$ evolves to $|\downarrow\downarrow e\rangle$,
while the other states recover their original states as follows,
\begin{eqnarray}
&&|\downarrow\downarrow g\rangle \longrightarrow |\downarrow\downarrow e\rangle,~
|\downarrow\uparrow g\rangle \longrightarrow |\downarrow\uparrow g\rangle,\nonumber\\
&&|\uparrow\downarrow g\rangle \longrightarrow |\uparrow\downarrow g\rangle,~
|\uparrow\uparrow g\rangle \longrightarrow  |\uparrow\uparrow g\rangle.
\end{eqnarray}
Consequently, if the control qubit state is $|\downarrow\downarrow\rangle$,
the target qubit flips between $|g\rangle$ and  $|e\rangle$,
while for the other control qubit states the target qubit remains in the original state.

\begin{figure}[t]
\vspace{10cm}
\includegraphics{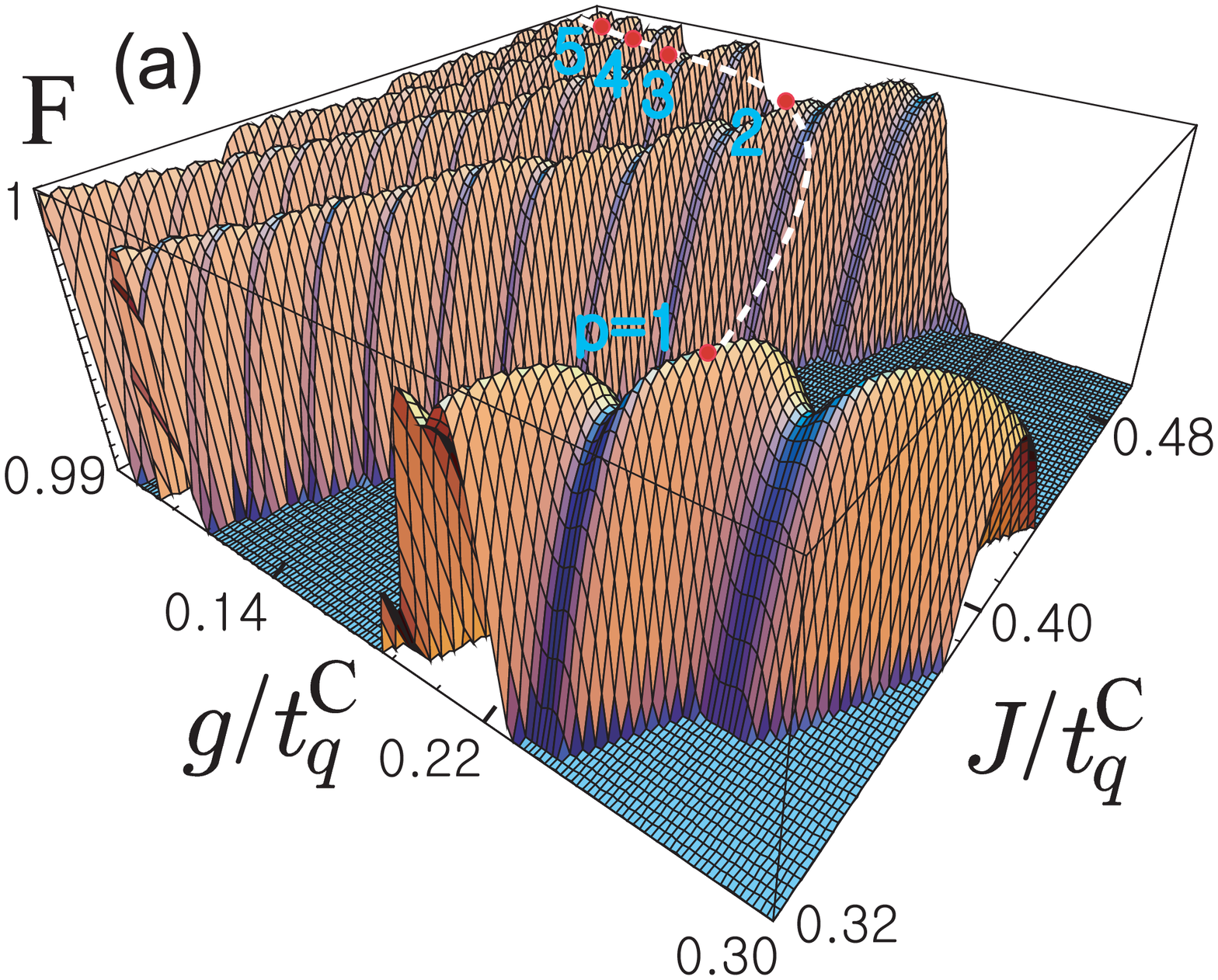}
\includegraphics{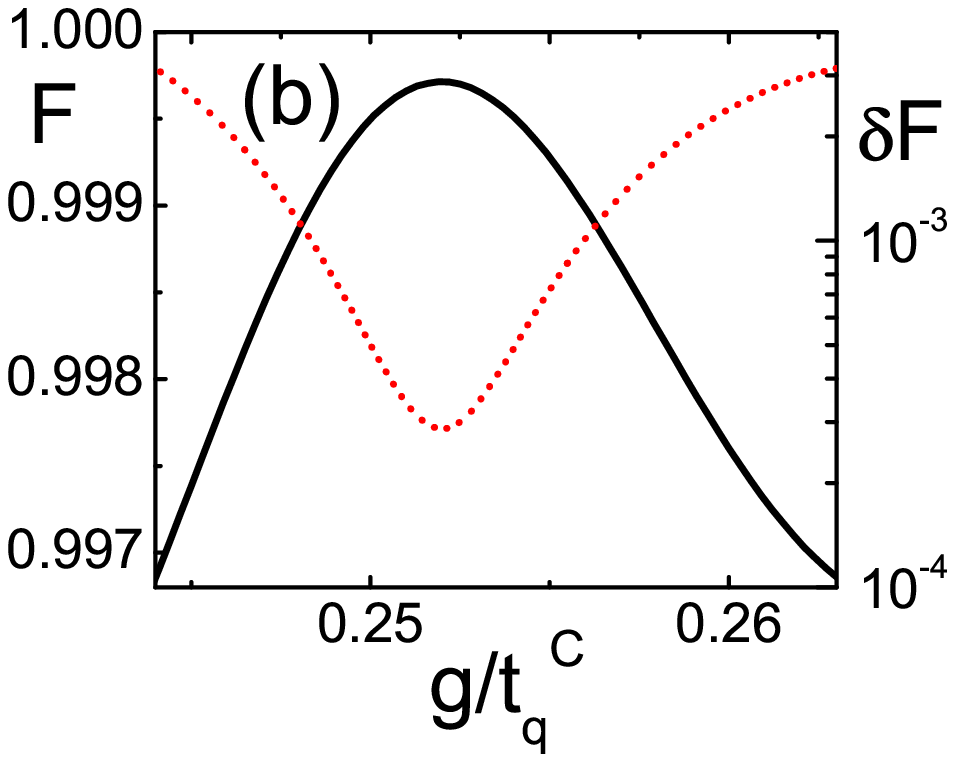}
\includegraphics{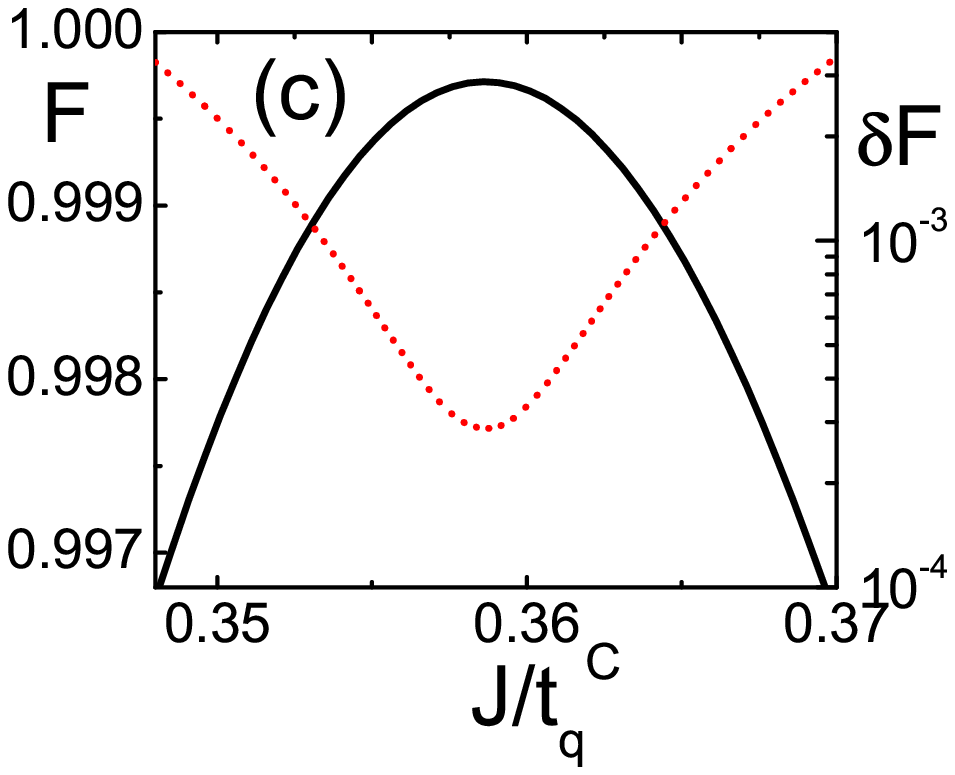}
\vspace{-1cm}
\caption{(Color online) (a) Fidelity $F$ in the plane of $(J,g)$ calculated numerically  with $q=3p$.
The peak points are shown in Table \ref{table}, which corresponds to the values
obtained analytically from Eqs. (\ref{condition}). (b) Cut view of (a) along the
$g$-axis and (c) $J$-axis for $(p,q)=(1,3)$(solid lines). Dotted lines show
the fidelity error $\delta F=1-F$ in log scale.}
\label{peak}
\end{figure}

We can observe that the parameter values obtained by the RWA well
describe the unitary evolutions for the Toffoli gate. The fidelity
of Toffoli gate can be calculated by $F={\rm Tr}(M(t)M_{\rm
Toffoli})/8$, where $M(t)$ is the truth table of Toffoli gate
calculated numerically. Figure  \ref{osc} (b) shows the fidelity
of the Toffoli gate for $(p,q)=(1,3)$ and (2,6). We can observe
that at $\Omega_0 t= (2k+1) \pi$ the single-step Toffoi gate is
achieved. The peak width of fidelity in Fig.  \ref{osc} (b) is
calculated as $\delta t \sim $  0.5 ns at $F=0.999$ with
$t^C_q$=1GHz, which is sufficiently wide for current state-of-the-art technology.

Figure \ref{peak} (a) shows the fidelity $F$ in the plane of $(J,g)$.
The peak points denote the local maxima of fidelity for $q=3p$,
which are listed in Table \ref{table}.
Since the RWA works well for weak coupling,
the RWA results  fit well with the numerical calculation
for small $g$ in Table \ref{table}.
Figure \ref{peak} (b) and (c) show the cut views of fidelity in (a) with
$(p,q)=(1,3)$ along the $g$-axis and $J$-axis, respectively, which pass the local maximum.
The peak width with $F=0.999$ is $\delta g/t^C_q \approx 0.008$ and
$\delta J/t^C_q \approx 0.01$. For $t^C_q=1-2$GHz, $\delta g$ and $\delta J$ are
of the order of 10MHz.

In Fig. \ref{peak} (b) and (c) the fidelity errors $\delta F=1-F$
are shown. The maximum fidelity (F=1) can be achieved by using the Jaynes-Cummings
Hamiltonian of Eq. (\ref{HRWA}) in the RWA.
In fact, however, our numerical calculation is based on the
semiclassical Hamiltonian of Eq. (\ref{Hone}), which includes
multi-photon processes and fast mode oscillations. As a result, due to these
imperfections  we have the fidelity error;
$\delta F \sim 3\times 10^{-4}$ which is 3 times larger than
$\delta F \sim 10^{-4}$ for the two-qubit oscillations.
\cite{comm} The 3-qubit Toffoli gate can be implemented by
decomposing it into a sequence of 10 single-qubit operations and 6
controlled-NOT operations. \cite{Barenco} Although the fidelity
error for the present three-qubit oscillation is larger compared
to two-qubit gate operation, our single-step implementation scheme
has an advantage in this respect.

\section{Decoherence Analysis}

The environment surrounding the qubit system is known to invoke a
decoherence in the qubit state. The environment can be described as a
reservoir of thermal bath.
We consider that our qubit of Eq. (\ref{HRWA}) is coupled with the reservoir.
If we trace out the reservoir degree of freedom from the total
density matrix, the time evolution of the reduced density matrix $\rho$
is given by the Master equation in the Born-Markov approximation, \cite{Lou,Kuang,Moya}
\begin{eqnarray}
\label{Master}
\frac{d\rho}{dt}=-\frac{i}{\hbar}[{\tilde {\cal H}}^{\rm I,RWA}_{|ss'\rangle},\rho]
-\frac{\gamma}{\hbar^2}[{\tilde {\cal H}}^{\rm I,RWA}_{|ss'\rangle},[{\tilde {\cal H}}^{\rm I,RWA}_{|ss'\rangle},\rho]],
\end{eqnarray}
where $\gamma$ depends on the temperature and the spectral
density of the reservoir.

Here, if we introduce a dressed state basis,
\begin{eqnarray}
\label{dress}
|+n\rangle&=&\cos\phi_{jn}|g,n+1\rangle+\sin\phi_{jn}|en\rangle, \nonumber\\
|-n\rangle&=&-\sin\phi_{jn}|g,n+1\rangle+\cos\phi_{jn}|en\rangle
\end{eqnarray}
with
\begin{eqnarray}
\tan\phi_{jn}\!=\!\frac{{\tilde g}_{jn}\sqrt{n+1}}{\hbar(\omega_j-\omega_0)\!\!+\!\!\sqrt{{\tilde g}^2_{jn}(n+1)\!+\!\hbar^2(\omega_j-\omega_0)^2}},
\end{eqnarray}
the qubit Hamiltonian of Eq. (\ref{HRWA}) can be diagonalized as
\begin{eqnarray}
&&{\tilde {\cal H}}^{\rm I,RWA}_{|ss'\rangle}|\pm n\rangle=E^{\pm}_{jn}|\pm\rangle, \\
&&E^{\pm}_{jn}=\hbar\omega\left(n+\frac12\right)\pm\frac{1}{2}\sqrt{{\tilde g}^2_{jn}(n+1)+\hbar^2(\omega_j-\omega_0)^2}.
\nonumber
\end{eqnarray}

In this dressed state basis the Master equation of Eq. (\ref{Master})
can be exactly solvable, providing the time evolution of the density matrix ~\cite{Gang}
\begin{eqnarray}
\frac{d}{dt}\rho^{pp'}_{nm}(t)\!=\!-\!\!\left[\frac{i}{\hbar}(E^p_{jn}\!-\!E^{p'}_{jm})
\!+\!\frac{\gamma}{\hbar^2}(E^p_{jn}\!-\!E^{p'}_{jm})^2\!\right]\!\! \rho^{pp'}_{nm}(t),\nonumber\\
\end{eqnarray}
where $p,p'\in\{+,-\}$ and
\begin{eqnarray}
\rho   &=&\rho^{++}_{nm}|+n\rangle\langle +m|+\rho^{+-}_{nm}|+n\rangle\langle -m|\nonumber\\
&&+\rho^{-+}_{nm}|-n\rangle\langle +m|+\rho^{--}_{nm}|-n\rangle\langle -m|.
\end{eqnarray}



Let us consider that initially the target qubit is at the ground state,
$|\psi(0)\rangle=|g,n+1\rangle$. The density matrix
$\rho(0)=|g,n+1\rangle\langle g,n+1|$ can be represented in the dressed state basis
by using Eq. (\ref{dress}). Then, the density matrix of  initial  state is given by
\begin{eqnarray}
\rho^{++}_{nn}(0)=\cos^2\phi_{jn}, ~~ \rho^{+-}_{nn}(0)=-\cos\phi_{jn}\sin\phi_{jn}, \nonumber\\
\rho^{-+}_{nn}(0)=-\sin\phi_{jn}\cos\phi_{jn}, ~~ \rho^{--}_{nn}(0)=\sin^2\phi_{jn}. ~~
\end{eqnarray}
By tracing out the density matrix over the photon field we obtain the density matrix of the qubit system
$\Upsilon=\sum_l\langle l|\rho|l \rangle$, where
$\Upsilon$ is a density matrix in the basis of $\{|e\rangle,|g\rangle\}$.
The expectation value of ground state population $Q_{ss'g}(t)=-{\rm Tr}[\Upsilon\sigma_z]$ is calculated as
\begin{eqnarray}
Q_{ss'{\rm g}}(t)=\cos^2 2\phi_{jn}+\sin^2 2\phi_{jn}\cos(\Omega_{jn}t)e^{-\gamma \Omega^2_{jn} t},
\end{eqnarray}
where $\Omega_{jn}\equiv (E^{+}_{jn}-E^{-}_{jn})/\hbar$.

In our scheme, for the control qubit state $|\downarrow\downarrow\rangle$ ($j=0$),
the target qubit states are resonant with the photon field.
For the resonant case the population expectation has been derived. \cite{Moya}
In our case  $Q_{\downarrow\downarrow{\rm g}}(t)$ is reduced to
\begin{eqnarray}
Q_{\downarrow\downarrow{\rm g}}(t)=\cos({\tilde g}_{0n}\sqrt{n+1}t/\hbar)e^{-\gamma {\tilde g}_{0n}^2(n+1) t/\hbar^2},
\end{eqnarray}
which is consistent with the previous result.

\begin{figure}[t]
\vspace{7.2cm}
\includegraphics{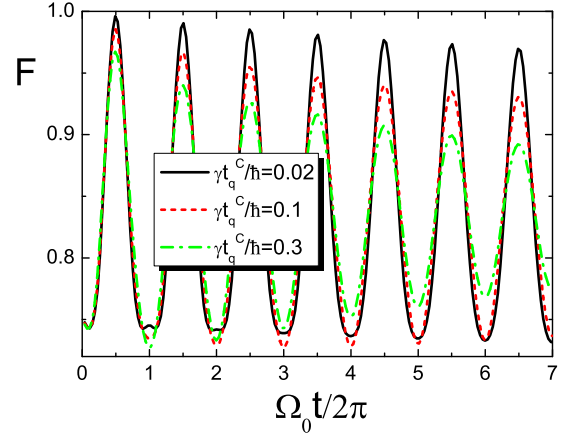}
\vspace{-0.8cm}
\caption{(Color online) Fidelity of Toffoli gate in the presence of the reservoir
for several qubit-reservoir coupling $\gamma$'s.
}
\label{damp}
\end{figure}

Since  we are considering the case $n=0$ in this study, we have
\begin{eqnarray}
Q_{ss'{\rm g}}(t)=\cos^2 2\phi_j+\sin^2 2\phi_j\cos(\Omega_jt)e^{-\gamma \Omega^2_j t}
\end{eqnarray}
with $\Omega_j=\Omega_{j0}$ and $\phi_j=\phi_{j0}$.
Here, we introduce the occupation probabilities
\begin{eqnarray}
\label{occ}
P_{ss'g}(t)\!=\!\frac{1\!+Q_{ss'{\rm g}}(t)}{2} ~~{\rm and}~~ P_{ss'e}(t)\!=\!1\!-\!P_{ss'g}(t).
\end{eqnarray}
For $\gamma=0$ these probabilities correspond to the numerical
result of  Fig. \ref{osc} (a), where the initial target qubit
state is the ground state.

The fidelity for the Toffoli gate is given by $F={\rm Tr}(M(t)M_{\rm Toffoli})/8$,
where the truth table matrix $M(t)$ has a block diagonal form
\begin{equation}
M(t)=M(t)_{|\uparrow\uparrow\rangle}\oplus M(t)_{|\uparrow\downarrow\rangle}\oplus
M(t)_{|\downarrow\uparrow\rangle}\oplus M(t)_{|\downarrow\downarrow\rangle}
\end{equation}
with
\begin{eqnarray}
M(t)_{|ss'\rangle}=\left(\matrix{P_{|ss'e\rangle\rightarrow |ss'e\rangle}(t) & P_{|ss'e\rangle\rightarrow |ss'g\rangle}(t) \cr
P_{|ss'g\rangle\rightarrow |ss'e\rangle}(t) & P_{|ss'g\rangle\rightarrow |ss'g\rangle}(t) }\right).
\end{eqnarray}
Here, if initially the qubit evolves from the ground (excited) state,
$P_{|ss'g(e)\rangle\rightarrow |ss'g(e)\rangle}(t)$ is the probability that
the qubit will occupy the ground (excited) state  at time $t$.  Further,
we have the relations that $P_{|ss'g\rangle\rightarrow |ss'g\rangle}(t)=
P_{|ss'e\rangle\rightarrow |ss'e\rangle}(t)
=1-P_{|ss'g\rangle\rightarrow |ss'e\rangle}(t)
=1-P_{|ss'e\rangle\rightarrow |ss'g\rangle}(t)$.
The probability $P_{|ss'g(e)\rangle\rightarrow |ss'g(e)\rangle}(t)$ is given by
the occupation probability of Eq. (\ref{occ}) as
\begin{eqnarray}
P_{|ss'g\rangle\rightarrow |ss'g\rangle}(t)=P_{|ss'e\rangle\rightarrow |ss'e\rangle}(t)=
\frac{1+Q_{ss'{\rm g}}(t)}{2}
\end{eqnarray}
and, thus, the fidelity is written by
\begin{eqnarray}
F=\frac18 [4+Q_{\uparrow\uparrow{\rm g}}(t)+Q_{\uparrow\downarrow{\rm g}}(t)
+Q_{\downarrow\uparrow{\rm g}}(t)-Q_{\downarrow\downarrow{\rm g}}(t)].
\end{eqnarray}
We show the damping of fidelity in Fig. \ref{damp} for several $\gamma$'s.
As shown in Fig. \ref{damp}, the maximum fidelity decays with the operation time.

\section{Summary}

We study a scheme for the single-step Toffoli gate for three  qubits
coupled by the Ising interaction.
The photon field is resonant with target qubit for a specific control qubit state
whereas it is off-resonant for the other control qubit states.
For the target qubit state with a resonant photon field, the qubit-photon system is
described by the Jaynes-Cummings model demonstrating a Rabi oscillation,
while for the other control qubit states it is described by a modified Jaynes-Cummings model.
We found that in the RWA the modified Jaynes-Cummings model is reduced to the usual
Jaynes-Cummings model with a renormalized coupling between qubit and photon.
The single-step Toffoli gate can be achieved for a $\pi/2$ rotation,
if the oscillation periods of target qubit
satisfy  commensurate conditions. The commensurate condition determines
the values of the qubit-photon coupling $g$ and the qubit-qubit coupling $J$
for achieving the Toffoli gate.
These values fit well with those obtained by numerical calculation for weak coupling $g$.
The fidelity is shown to be high and the peak width of the fidelity
to be wide enough for implementing the Toffoli gate.
The fidelity error is shown to be small, and thus our scheme has advantages over
the decomposition scheme. The decoherence effect from the environment is discussed.

\begin{center}
{\bf ACKNOWLEDGMENTS}
\end{center}

This work was supported by the NSFC under Grant No. 10874252. This
research was supported by Basic Science Research Program through
the National Research Foundation of Korea (NRF) funded by the
Ministry of Education, Science and Technology (2011-0023467; MDK).

\appendix
\section{}

According to the Baker-Campbell-Hausdorff formula
one of the terms in Hamiltonian of Eq. (\ref{tHI}), for example, is
\begin{eqnarray}
&&a(t)e^{-\gamma_j(a^\dagger(t)-a(t))}\sigma_-(t) \\
&&=a(t)\sum_{r,s}\frac{(-\gamma_ja^\dagger(t))^r}{r!}
\frac{(\gamma_ja(t))^s}{s!}e^{-\gamma^2_j/2}\sigma_-(t), \nonumber
\end{eqnarray}
where
\begin{eqnarray}
\gamma_j\equiv \frac{\beta_jg}{\hbar\omega_0}.
\end{eqnarray}
In this case, by using Eq. (\ref{sigma}) the time dependent part of the term is given by
$e^{i(r-s-1)\omega t-i\omega_0 t}$.
In the RWA  fast oscillating terms are neglected.
Hence, in order for the term to survive in the RWA when $\omega=\omega_0$, it should be that $r-s=2$.
This  means that for this term
we need to evaluate $\langle n+2|e^{-\gamma_j(a^\dagger(t)-a(t))}|n\rangle$ and then
the only non-vanishing matrix element is
\begin{eqnarray}
\langle g,n+1|a(t)e^{-\gamma_j(a^\dagger(t)-a(t))} \sigma_-(t)|e,n\rangle.
\label{nonv1}
\end{eqnarray}
Here, $|g\rangle$ and $|e\rangle$ denote the qubit states,
and $|n\rangle$ and $|n+1\rangle$ the photon number states.
For another term $a^\dagger(t)e^{-\gamma_j(a^\dagger(t)-a(t))}\sigma_-(t)$,
it should be that $r=s$, and  we need to evaluate
$\langle n|e^{-\gamma_j(a^\dagger(t)-a(t))}|n\rangle$.
Then, the only non-vanishing element is
\begin{eqnarray}
\langle g,n+1|a^\dagger(t)e^{-\gamma_j(a^\dagger(t)-a(t))}
\sigma_-(t)|e,n\rangle.
\label{nonv2}
\end{eqnarray}
The terms with $\sigma_+(t)$ and $\sigma_z$ in the Hamiltonian of Eq. (\ref{tHI})
can also be evaluated in a similar manner.

These matrix elements can be evaluated by the formula for
the displaced number state; \cite{Vogel} for $n\leq m$,
\begin{eqnarray}
\langle n|e^{\gamma a^\dagger-\gamma^* a}|m\rangle
=(-\gamma^*)^{m-n}\sqrt{\frac{n!}{m!}}L^{(m-n)}_n(|\gamma|^2)e^{-|\gamma|^2/2},\nonumber\\
\end{eqnarray}
and for $n>m$, $\langle n|e^{\gamma a^\dagger-\gamma^* a}|m\rangle
=\langle m|e^{-\gamma a^\dagger+\gamma^* a}|n\rangle^*$,
where $L^k_n(x)$ is the associated Laguerre polynomial.

By summing up all the contributions we have
\begin{eqnarray}
&&\langle g,n+1|\left(a(t)+a^\dagger(t)-\gamma_j\sigma_z\right)e^{-\gamma_j(a^\dagger(t)-a(t))}
\sigma_-(t)|e,n\rangle \nonumber\\
&&=\left(\sqrt{n+1}L^0_n(\gamma_j^2)+\frac{\gamma_j^2}{\sqrt{n+1}}[L^2_n(\gamma_j^2)-L^1_n(\gamma_j^2)]\right)
\nonumber\\ && ~~~\times e^{-\gamma_j^2/2}\equiv F_n(\gamma_j), \\
&&\langle e,n|\left(a(t)+a^\dagger(t)-\gamma_j\sigma_z\right)e^{\gamma_j(a^\dagger(t)-a(t))}
\sigma_+(t)|g,n+1\rangle \nonumber\\
&&=\left(\sqrt{n+1}L^0_{n+1}(\gamma_j^2)+\frac{\gamma_j^2}{\sqrt{n+1}}[L^2_{n-1}(\gamma_j^2)+L^1_n(\gamma_j^2)]\right)
\nonumber\\ &&~~~\times e^{-\gamma_j^2/2}\equiv G_n(\gamma_j).
\end{eqnarray}

\end{document}